\documentclass[
reprint,
superscriptaddress,
 amsmath,amssymb,
 aps,
pra,
]{revtex4-2}

\usepackage{graphicx}
\usepackage{dcolumn}
\usepackage{bm}
\usepackage[utf8]{inputenc}
\usepackage[T1]{fontenc}     
\usepackage{lmodern}    

\begin{document}

\preprint{APS/123-QED}

\title{Noise-robust classification of single-shot electron spin readouts\\
using a deep neural network}

\author{Yuta Matsumoto}
 \email{ymatsumoto21@sanken.osaka-u.ac.jp}
\affiliation{The Institute of Scientific and Industrial Research, Osaka University,8-1 Mihogaoka, Ibaraki, Osaka 567-0047, Japan}
\author{Takafumi Fujita}
\affiliation{The Institute of Scientific and Industrial Research, Osaka University,8-1 Mihogaoka, Ibaraki, Osaka 567-0047, Japan}
\affiliation{Artificial Intelligence Research Center, The Institute of Scientific and Industrial Research, Osaka University, 8-1 Mihogaoka, Ibaraki, Osaka 567-0047, Japan}
\affiliation{Center for Spintronics Research Network, Graduate School of Engineering Science, Osaka University, 1-3 Machikaneyama, Toyonaka, Osaka 560-8531, Japan}
\author{Arne Ludwig}
\affiliation{Lehrstuhl f{\"u}r Angewandte Festk{\"o}rperphysik, Ruhr-Universit{\"a}t Bochum, Universit{\"a}tsstra{\ss}e 150, Geb{\"a}ude NB, D-44780, Germany}
\author{Andreas D. Wieck}
\affiliation{Lehrstuhl f{\"u}r Angewandte Festk{\"o}rperphysik, Ruhr-Universit{\"a}t Bochum, Universit{\"a}tsstra{\ss}e 150, Geb{\"a}ude NB, D-44780, Germany}

\author{Kazunori Komatani}
\affiliation{The Institute of Scientific and Industrial Research, Osaka University,8-1 Mihogaoka, Ibaraki, Osaka 567-0047, Japan}
\affiliation{Artificial Intelligence Research Center, The Institute of Scientific and Industrial
Research, Osaka University, 8-1 Mihogaoka, Ibaraki, Osaka 567-0047, Japan}
\author{Akira Oiwa}
\affiliation{The Institute of Scientific and Industrial Research, Osaka University,8-1 Mihogaoka, Ibaraki, Osaka 567-0047, Japan}
\affiliation{Center for Spintronics Research Network, Graduate School of Engineering Science, Osaka University, 1-3 Machikaneyama, Toyonaka, Osaka 560-8531, Japan}
\affiliation{Center for Quantum Information and Quantum Biology, Institute for Open and Transdisciplinary Research Initiatives, Osaka University, 1-2 Machikaneyama, Tokyonaka, Osaka 560-0043, Japan}

\begin{abstract}
Single-shot readout of charge and spin states by charge sensors such as quantum point contacts and quantum dots are essential technologies for the operation of semiconductor spin qubits. The fidelity of the single-shot readout depends both on experimental conditions such as signal-to-noise ratio, system temperature and numerical parameters such as threshold values. Accurate charge sensing schemes that are robust under noisy environments are indispensable for developing a scalable fault-tolerant quantum computation architecture. In this study, we present a novel single-shot readout classification method that is robust to noises using a deep neural network (DNN). Importantly, the DNN classifier is automatically configured for spin-up and spin-down signals in any noise environment by tuning the trainable parameters using the datasets of charge transition signals experimentally obtained at a charging line. Moreover, we verify that our DNN classification is robust under noisy environment in comparison to the two conventional classification methods used for charge and spin state measurements in various quantum dot experiments.

\end{abstract}
\maketitle
\flushbottom

%
%
\thispagestyle{empty}

\section{Introduction}
Electron spins in semiconductor quantum dots (QDs) are a promising candidate for quantum bits (qubits), owing to their potential scalability\cite{Vander,pet}, for fault-tolerant quantum computing. Spin qubits in QDs are measured using single-shot readouts via a spin-to-charge conversion technique in which spin states are distinguished by the presence of a charge transition event in the sensor signals\cite{elzer}. Such sensors are usually located close to QDs and are capacitively coupled to QDs. For the scalability of universal quantum computation, a high fidelity ($\geq \mathrm{99\%}$), is required for all processes including preparation, control and measurement of single qubits to implement the surface code error correction protocol\cite{qcomp,qcomp2}. Qubits must be detected using a limited number of dedicated sensors in a scaled-up system, but having a small capacitance limits the sensor sensitivity. To overcome this issue, qubit readouts in large-scale dense qubit arrays are likely performed using dispersive gate sensing\cite{disper,disper2}, which can be flexibly placed without carrier reservoirs and supports compactness of the device architecture; however, it has an inherently low sensitivity. The single spin readout fidelity can be optimized by adjusting experimental parameters such as readout bandwidth, spin relaxation rates, filter rates, data sampling rates and electron tunnel rates\cite{Keith_2019}. 

Noise sources harm the signal detection accuracy and limit the detection bandwidth and filtering rates, thereby affecting aforementioned parameters. The signal-to-noise ratio (SNR) is increased through both the hardware and the parameter tuning of classification algorithms, to enable faster detection of the small signals which were limited by the untunable physical parameters\cite{fastreadout}. Low-dimensional quantum devices experience several noises such as amplifier noise, shot noise and Johnson-Nyquist noise. These appear as Gaussian noise in the sensor signals\cite{gaussian}. The 1/$f$ noise caused by charge fluctuations in the vicinity of the sensor and qubit appears as voltage drift with a lower frequency than the data sampling rates (drift noise)\cite{drift}. Other potential noise
sources are the spike-like noise due to instability in voltage sources, and the interference from the alternative current (AC) power supply. In the case of such noise sources with a small SNR, it is hard to find an appropriate physical or numerical filter that provides high-fidelity single-shot readouts with a large bandwidth. Therefore, a precise post-processing of such noisy single-shot signals is vital for a fast high fidelity single-shot readout in any large-scale dense QD array.
\begin{figure*}
\centering
\includegraphics[width=18cm]{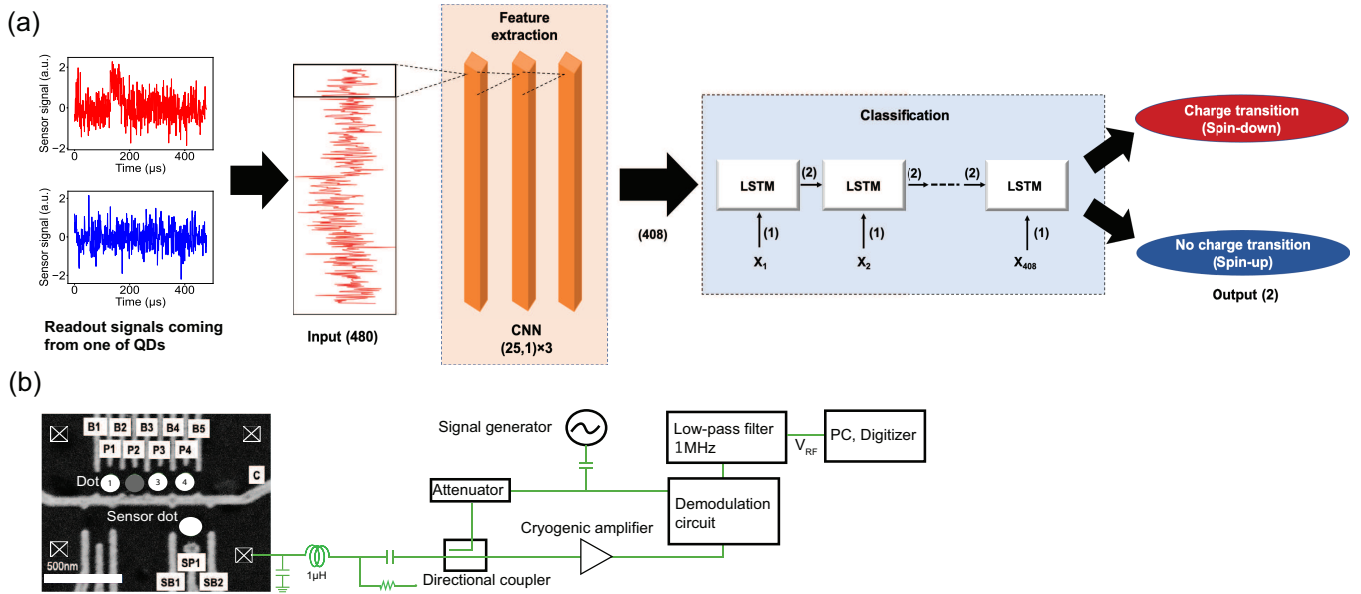}
\caption{\label{fig:wide}(a) Scheme of the proposed DNN classification method, where (n) indicates the number of dimensions. The red and blue traces are the examples of sensor signals with (red) or without a charge transition event (blue), respectively. Input signals for the DNN classifier are the measured voltage value of 480 $\mu$s RF signals per $\mu$s. The CNN extracts the features of charge transition events from the measured single-shot readout signal. The LSTM classifies the extracted features in the CNN ($X_1$\textasciitilde$X_{408}$) whether a charge transition is present or not. (b) The SEM image of a gate-defined GaAs multiple QD device that is nominally identical to the one used for measurements (see methods in detail). The solid circles indicate QDs. The white squares marked with a cross show ohmic contacts. The RF reflectometry circuit for charge sensing was connected to one of the ohmic contacts. In all measurements, the RF signals are filtered by a 1MHz low-pass filter.}
\end{figure*}
Although several post-processing techniques have been proposed to reduce the spin readout error\cite{wavelet,likelihoodest,optimalfilter}, no studies have applied machine learning (ML), specifically the deep neural network (DNN), to distinguish real-time charge transition events that are obscured by noises in quantum devices. We note that in the field of QDs, a few studies have used DNNs for the recognition of two-dimensional charge stability diagrams of multiple QDs\cite{machineauto,autosim,auto3}.

\begin{figure*}
\centering 
\includegraphics[width=18cm]{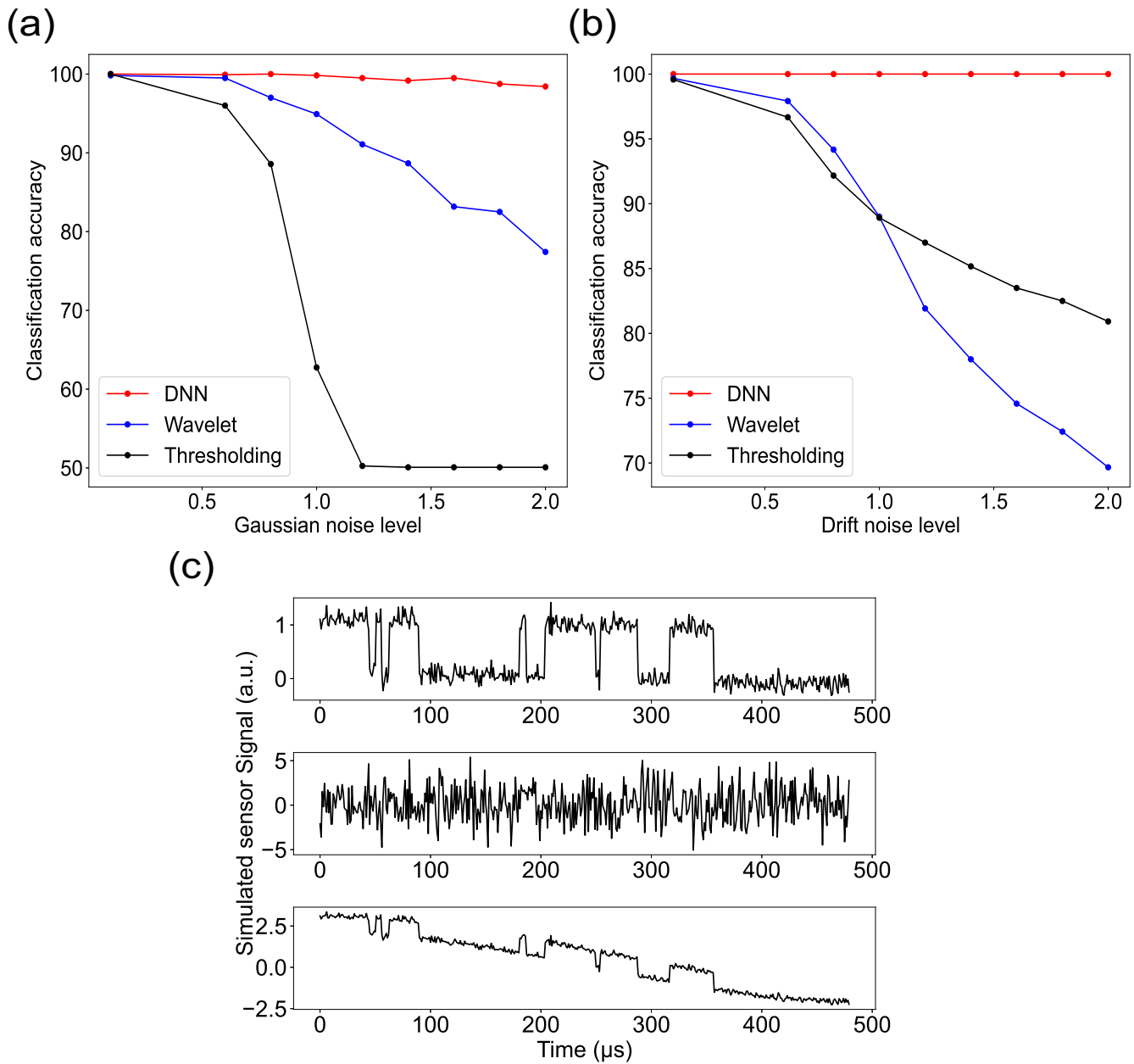}
\caption{\label{fig:wide} Comparisons of classification accuracies among the DNN classification, the wavelet method, and thresholding as a function of (a) Gaussian noise level and (b) drift noise level added to the artificial signals that are generated by the Markov chain model. (c) Simulated real-time sensor signals including charge transition events with Gaussian noise level of 0.1 (top), Gaussian noise level of 2.0 (middle), and drift noise level of 2.0 (also added Gaussian noise level of 0.1 to imitate the sensor signals) (bottom).}
\end{figure*}
In this study, to realize high fidelity spin measurements in QDs with a device-limited small SNR under harsh noise environments, we present a new signal classification method based on the DNN to classify single-shot electron spin readout signals measured by a charge sensor in gate-defined QDs. We first evaluate the proposed DNN classification with simulated charge transition events in sensor signals with various noise environments. Next, we show the training procedure to apply the DNN classification to the real QD devices and evaluate the classification accuracy of charge transition events occurring in GaAs-based lateral QDs. Finally, we demonstrate the precise electron spin single-shot measurements using the trained DNN classifier in a noisy environment.
\section{Results}
\subsection{Neural network architecture}
First of all, we describe the architecture of our DNN classifier. The single-shot classification procedure of our DNN is shown in Fig. 1(a). The DNN consists of a convolutional neural network (CNN) for feature extraction and the long short-term memory (LSTM) for classification. The dimension of the input is 480. The outputs are two classified indices, either including the charge transitions or not. The filter size of the CNNs is (25,1), each with a stride of 1. The total numbers of parameters in the LSTM and in the DNN classifier are 32, and 110, respectively. We aim to classify each voltage trace and output as these two labels using the DNN classifier. The original data is a time trace of voltage outputs $V_{RF}$ of a demodulated radio-frequency (RF) signal, which is reflected from the charge sensor. Such a time trace of a fixed acquisition length represents a single spin detection by observing whether a charge transition is present (labeled as spin-down) or not (labeled as spin-up). We choose the measured voltage value of 480 $\mu$s RF signal per $\mu$s, which stochastically includes the features of the charge transition event, as the input for the neural networks. Before the DNN classifying step, the mean value of the signals measured at a condition without charge transition events is subtracted from the raw data to standardize the baseline of the input signals to zero. We input each pre-processed RF voltage value into each input neuron per trace. The input signals go through three CNN layers, in which each neuron outputs spatially correlated information for the data points inside a given filter size to extract voltage shift features associated with charge transition events. In the next step, the signals are processed by the LSTM layer, which learns from the correlations among different time series. This increases the robustness to low frequency noise compared with using only CNNs. The input vector size of the LSTM is 1 and the LSTM outputs the probabilities for classifying the single-shot signal with or without a charge transition by the softmax activation function. The decision of detecting a transition signal classifies the trace as either spin-up or spin-down. Although we changed the number of input neurons according to sequence lengths of a single-shot data, the total number of parameters in the DNN classifier is constant and the DNN classifier maintained high classification accuracy. Deep learning can flexibly tune the trainable parameters through a simple procedure, depending on the noises and sensor sensitivity for event detections. This is a remarkable advantage of the DNN classification; it automatically builds an appropriate algorithm for each spin qubit, which has a small SNR. We defined SNR as 20$\rm{log_{10}}$($A_{\rm{signal}}$/$A_{\rm{noise}})$ (dB), where $A_{\rm{signal}}$ and $A_{\rm{noise}}$ are root-mean-square values of the signal and noise, respectively. 

\begin{figure*}
\centering
\includegraphics[width=19cm]{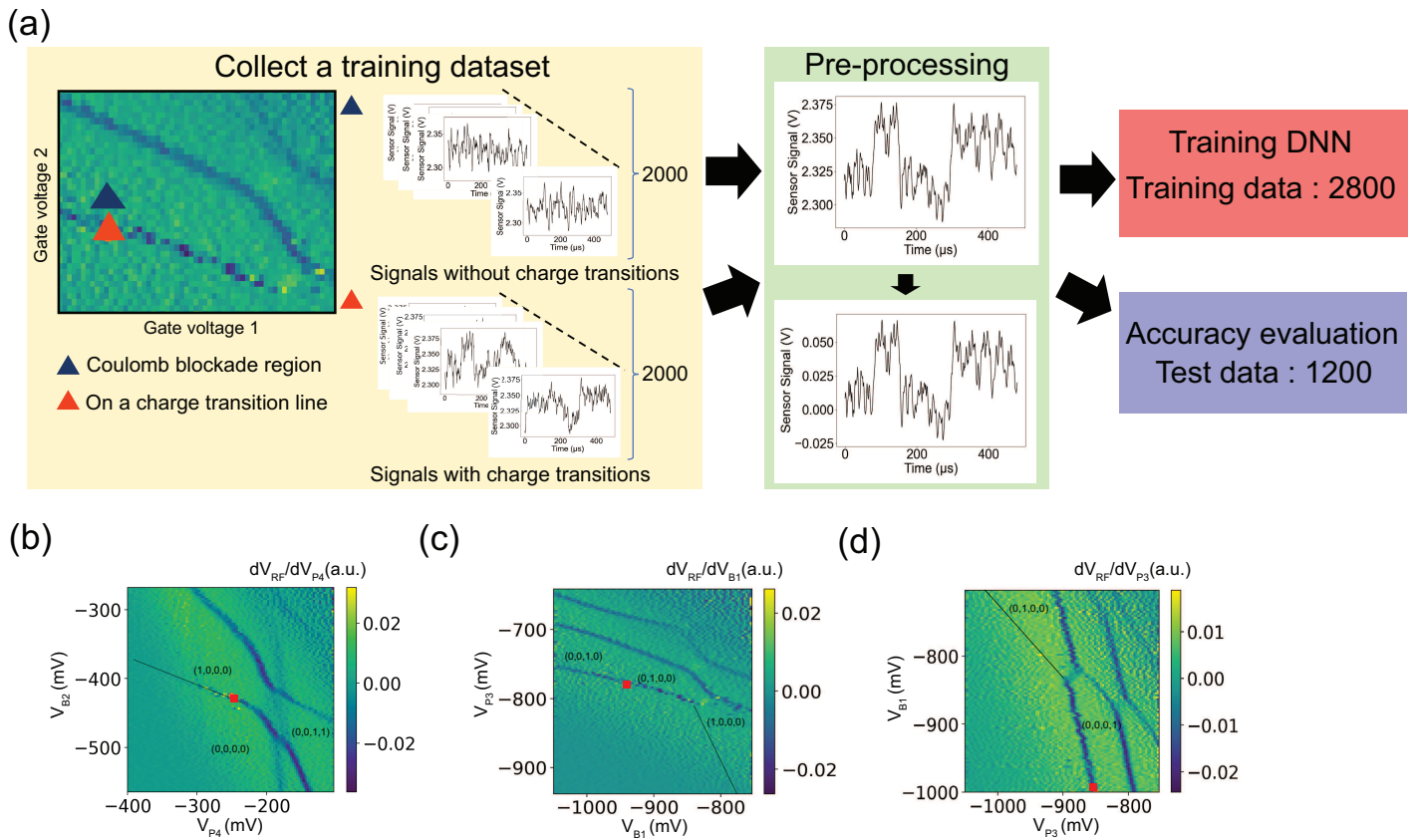}
\caption{\label{fig:wide}(a) Schematic of the procedure for collecting and pre-processing the dataset for training. Each single-shot data is pre-processed by subtracting the mean value of the signals measured at a condition without charge transition events to remove the offset amplitude of charge transition signals. Then, 2800 pre-processed signal sequences, which have the standardized voltage shift features associated with charge transition events and noises, are used to train the DNN classifier. Next, 1200 signal sequences are used to evaluate the classification accuracy. The stability diagrams are used to collect the sequence data in Dot1(b), Dot3(c), and Dot4(d). The red squares show the position for collecting the data with charge transition events.}
\end{figure*}
\subsection{Noise-robustness evaluation}
We evaluated the robustness of the DNN classifier against Gaussian noise and drift noise using simulated signals. We add the simulated Gaussian noise and drift noise on simulated signals to imitate the experimental noisy environments. We defined the noise level as the relative linear amplitude of noise to that of the charge event signal. We compared the results of the DNN classifications with two standard numerical analysis methods often used in gate-defined QDs for identifying charge transition events\cite{elzer,wavelet}: threshold-based classification (thresholding) and wavelet transformation classification (the wavelet method). Figure 2(a) shows the classification accuracies of the DNN classification, the wavelet method, and thresholding as a function of the Gaussian noise level. The classification accuracy is defined as the fraction of times when the output label matches the preassigned label of the simulated signals. We investigated the accuracy of thresholding with all the possible threshold values for each noise level and then chose the threshold values to locally maximize accuracy. As the noise level increased above 0.5, the accuracies for the wavelet method and thresholding gradually decreased. The accuracy for the DNN classification, however, did not drop significantly even when the noise level reached 2.0. This indicates that the DNN classification is more robust than conventional methods under the presence of certain Gaussian noise. Figure 2(b) shows the classification accuracies for the three methods as a function of the drift noise level. With increasing drift noise level, accuracy decreased for thresholding and the wavelet, whereas the DNN classification correctly classified all 1200 samples in the test dataset. These results indicate that the DNN classifier builds a numerical filter appropriate for different types of noise with different amplitude by tuning the trainable parameters.

Examples of the simulated signal are shown in Fig. 2(c). The top panel shows the almost ideal sensor signal that is accurately identified by all three classification methods. The middle shows the signal with the Gaussian noise (noise level of 2.0). The bottom shows the signal with drift noise (noise level of 2.0). Although the charge transition events were buried in Gaussian noise with noise level of 2.0 (the middle panel of Fig. 2(c)), the DNN classifier identified the transition events with relatively high accuracy in this harsh noisy situation.

\subsection{Training procedure and accuracy evaluation on QD devices}
Next, we demonstrated the advantage of the DNN classification using the experimental dataset, which was obtained from the sensor QD adjacent to a GaAs quadruple QD. The procedure to learn actual charge transition events is quite simple. As depicted in Fig. 3(a), it consists of three steps: (1) dataset collection, (2) pre-processing of the dataset (standardization), and (3) training the DNN classifier. 

In step (1), we collected 2000 charge signals on the charge transition line for electron number N changing between 0 and 1 as down-spin data and 2000 signals off the charge transition line as up-spin data. We assumed that a signal on the transition line contains at least one charge transition event if the readout time is longer than approximately fivefold tunnel time to the lead. This assumption was validated by the Markov chain model simulation. We were able to recognize the charge transition lines of each QD in the stability diagrams as shown in Figs. 3(b), (c), and (d). The dataset for training and analysis was collected for training and analyses at a transition line. In step (2), we subtracted the mean value of the signals measured at a condition without charge transition events from each single-shot data. Then, the offset amplitude of the charge transition signal was removed. Finally, in step (3), we trained the networks with the dataset that was prepared in steps (1) and (2). The data were sorted randomly into two datasets, 2800 data points to train the networks and 1200 data points to evaluate classification accuracy.
\begin{table*}
\centering
\caption{\label{tab:table3}Classification accuracy for DNN classification and thresholding for the same experimental charge sensor signals. The threshold value was optimized to minimize the error. The SNR in Dot1, Dot3, and Dot4 were 7.41 $\mathrm{dB}$, 13.5 $\mathrm{dB}$, and 11.5 $\mathrm{dB}$, respectively.}
\begin{tabular}{ccccc}
\hline
 Classification method&Accuracy to Dot1&Dot3&Dot4\\ \hline
 DNN&$97.8$\% & $95.3$\% &$96.1$\% \\
 Thresholding&$81.5$\% & $94.1$\% & $95.0$\%\\\hline
\end{tabular}
\end{table*}
Using the trained DNN classifier, we evaluate the classification accuracy of charge tunneling events for QDs with different SNRs under various noise environments in a QD array. Table 1 shows the accuracies in each QD for the DNN classification and thresholding. Similar to the previous analysis, the classification accuracy was defined as the fraction of times when the output state corresponds to the preassigned label in step (1). We achieved over 95\% accuracy in all the QDs using the DNN classification, regardless of the distance between the target QD and the sensor and, thus, the difference in the local noise environments. The accuracy of the thresholding in Dot1 located far from the charge sensor decreased to 81.5\%, which was lower than other QDs and worse than the accuracy expected from the SNR. The lower accuracy in Dot1 was mainly due to the presence of the spike-like noise caused by the instability of measurement equipments. It does not apppear in the SNR because it irregularly appears as a short time pulse signal. Since the accuracy in QD1 was drastically restored by using the DNN classification, this indicates that deep learning readily finds the appropriate numerical filter also for such spike-like noise without prior knowledge of the noise type. We note that the accuracy of the DNN classification was mainly limited  by wrong labelling of the training dataset in the experiment. A few signal obtained in the Coulomb blockade region accidentally contained charge transition events caused by thermal excitation and a few signal obtained on the charge transition line under certain conditions of tunnel rates, detection band width and readout rate did not contain any charge transition events. These errors can be easily solved with lower electron temperture and an appropriate device tuning.
\begin{figure*}
\centering
\includegraphics[width=16cm]{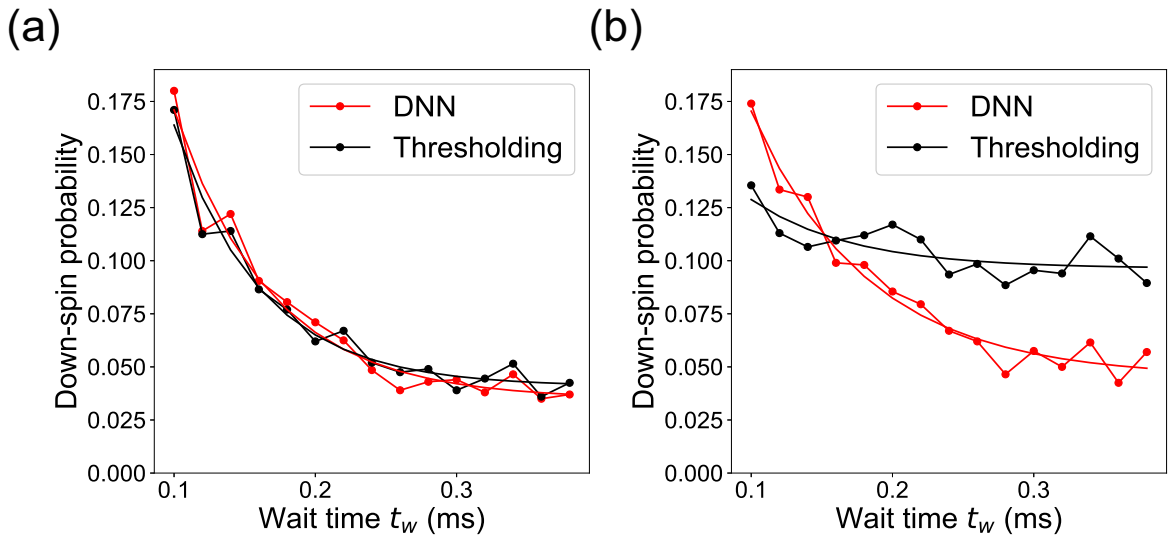}
\caption{\label{fig:wide} Spin relaxation measurement results using the DNN classification and thresholding in Dot4. In the DNN classification, we trained the network following the procedure as shown in Fig. 3(a) in Dot4. (a) Down-spin probability versus wait time $t_w$, out of a total 2000 traces taken for each waiting time for both the DNN classification and thresholding. The solid lines are the fit to an exponential decay curve (see details in the main text). The threshold value is optimized to maximize the visibility of decrease in the count of down-spin. (b) The down-spin probability versus wait time $t_w$ in the presence of artificially added Gaussian noise.}
\end{figure*}
\subsection{Spin state estimation}
Finally, to validate the DNN classification for extracting physical quantities, that do not change depending on the classification methods, we performed spin relaxation time ($T_1$) measurements using the DNN classification and thresholding in Dot4. We use the energy
selective readout scheme for spin-to-charge conversion\cite{elzer}. In this conversion method, only the electron in a spin excited state can tunnel off the dot. That is, an electron tunneling from a spin ground state is forbidden. Since the spin excited state is a spin-down state for electrons in GaAs QDs, we determine the spin-down probability by counting the number of data with a charge transition in the repeated single-shot measurements. Therefore, the classification accuracy of the charge transition events is crucial for the high-fidelity spin state readout. Figure 4(a) shows the down-spin probability estimated using the DNN classification and thresholding as a function of the wait time $t_w$, which is the duration for injecting electron spin and waiting spin relaxation in the QD. We used the trained network with the dataset obtained from Dot4. The red and black circles show the results of the DNN classification and thresholding, respectively. The relaxation times $T_1$ were obtained from the fitting of the exponential curve A $\times$ exp(-$t_w$/$T_1$) + B, where A and B are the fitting parameters. The relaxation times estimated by the DNN classification were within an error range of those obtained by thresholding. Thus, the advantage of the DNN classification was limited when the SNR was high enough (11.5 dB) for thresholding. The DNN classification is most powerful under noisy environments.

Therefore, to further investigate robustness to noise, we artificially added Gaussian noise to the single-shot data to create a noisy environment, which reduced the SNR to -2.51 dB (total noise level of 1.34). Figure 4(b) shows the estimated down-spin probability with artificially added Gaussian noise. A large decrease in the down-spin probability amplitude was observed for thresholding in the noisy environment, whereas the amplitude was almost unchanged for the DNN classification. The decrease in the probability amplitude leads to a large error in the estimated spin relaxation time as shown in Table 2. The error of $T_1$ significantly increased only for thresholding and not the DNN classification. Thus, the DNN classification provides more precise estimations of the spin relaxation times than thresholding, indicating the robustness of the DNN classification in a noisy environment for spin state identification. We note that the offset B, which implies the probability of the wrong outcome, “up-spin state” for down-spin state readout, increased for thresholding in the noisy environment, which also validates the robustness of the DNN classification for the spin state.

\section{Discussion}
We proposed and demonstrated a new method based on a DNN to classify noisy single-shot electron charge and spin readout signals. This DNN classification method flexibly builds numerical filters regardless of the noise type and outperformed two conventional numerical classifications. The performance of the charge tunneling event identification in QDs was experimentally improved using DNNs trained on measured datasets. To verify its applicability to spin state readouts, we showed that the DNN classification estimated the spin relaxation time more precisely than thresholding, especially in a noisy environment. This suggests that the DNN offers robust classifications of the charge sensor signals measured under noisy environments, allowing accurate extraction of various physical quantities in QDs. The accuracy of the DNN classifier can be further increased with a more sophisticated neural network architecture. This work validates the use of ML in building an appropriate classification algorithm to measure quantum devices and opens a route for future applications such as fast spin readout\cite{rapidreadout}, high temperature qubit operations\cite{1K,1K2} and dispersive gate readout\cite{disper,disper2}, where the SNR is generally small.
\begin{table*}
\centering
\caption{\label{tab:table3}$T_1$ times and fitting parameters extracted from spin relaxation measurements using DNN classification or thresholding.}
\begin{tabular}{ccccc}
\hline
 Classification method&$T_1$&Amplitude (A) &Offset (B) \\ \hline
 DNN& 68.19 $\pm$ 10.14 ($\mu$s) & 0.59 $\pm$ 0.13 &0.03 $\pm$ 0.01 \\
 Thresholding& 61.63 $\pm$ 8.35 ($\mu$s) & 0.62 $\pm$ 0.14 & 0.04 $\pm$ 0.00 \\
 DNN in a noisy environment& 82.43 $\pm$ 12.47 ($\mu$s) & 0.42 $\pm$ 0.08 & 0.05 $\pm$ 0.01 \\
 Thresholding in a noisy environment& 71.16 $\pm$ 43.77 ($\mu$s) & 0.13 $\pm$ 0.12 & 0.10 $\pm$ 0.01\\\hline
\end{tabular}
\end{table*}
\section{Methods}
\subsection{Device and measurements}
The quadruple quantum dot array is formed electrostatically in a two dimensional electron gas which is 90 nm below the surface of a GaAs/AlGaAs heterostructure (see Figure 1(b)). To operate the QDs, the center gate (C) separates the QD array and the sensor QD, the barrier gates (B1, B2, B3, B4, B5) tune the tunnel rates to the reservoir or the interdot, and the plunger gates (P1, P2, P3, P4) tune the chemical potential of individual QDs. The sensor dot gates (SB1, SB2, SP1) form a conventional sensitive QD sensor. The charge sensor conductivity, which is measured using RF reflectometry operating at a bandwidth up to 15 MHz, reflects the electron number in a QD array. All measurements were performed under an in-plane magnetic field of 3.2 T that enables clear distinction of spin-up and spin-down states. The device was cooled to a base temperature of 10 mK in a dilution refrigerator. 
\subsection{Simulated signal creation}
The simulated signals used to evaluate the robustness of the DNN classification are created by the Markov chain model
. We create 2000 signals containing charge transition events with a tunneling time of 33 $\mu$s (charge transitions
dataset) and 2000 signals without the transition events (no charge transition dataset). All the signals have a total length of 480 $\mu$s. The data are then sorted randomly into two datasets with 2800 signals used to train the DNN and 1200 signals used for classifier evaluation. We
choose a 1 kHz sinusoidal wave, which is slower than the measurement time (480 $\mu$s), as the simulated drift noise.
\subsection{Experimental Training Data Acquisition}
To use the DNN classification in real measurements of gate-defined QDs, we propose and demonstrate a practical
procedure to collect the enormous amount of labeled data required to train the DNN classifier for each sensor configuration by
setting the state on a charge transition line where sequential charge tunneling independent of the spin occurs. These real-time charge signals with charge transition events mimic the discriminator signals used to identify spin states. We perform acquisition of the datasets for training and single-shot spin readout measurements in a gate-defined QD array formed at a GaAs/AlGaAs heterointerface. 

\section*{Acknowledgements}
This work was supported by Grants-in-Aid for Scientific Research S (17H06120), JST CREST (JPMJCR15N2), the Asahi Glass Foundation, and  Dynamic Alliance for Open Innovation Bridging Human, Environment and Materials. A.L. and A.D.W. acknowledge gratefully support of DFG-TRR160 and  BMBF - Q.Link.X  16KIS0867.


\bibliography{main}

\end{document}